\documentclass[11pt,preprint,onecolumn,aps,longbibliography]{revtex4-1}
\usepackage{setspace}
\usepackage{graphicx}
\usepackage{amsmath} % The math mode package
\usepackage{mathtools}
\usepackage{amsfonts}
\usepackage{bbm}
\usepackage{braket}
\usepackage{caption}
\usepackage{graphicx}
\usepackage{float}
\usepackage{caption}
\usepackage{subcaption}
\usepackage{url}
\usepackage{color}
\usepackage[ruled]{algorithm2e}

\usepackage{amssymb}
\usepackage{cancel}
\usepackage{upgreek}
\usepackage{tabularx}
\usepackage{color}
\usepackage{units}
\usepackage{hhline}
\usepackage{textfit} %\scaletoheight{0.05cm}{very small text}
\usepackage{slashed}
\graphicspath{ {images/} }
\usepackage{graphicx}% Include figure files
\usepackage{dcolumn}% Align table columns on decimal point
\usepackage{bm}% bold math
\usepackage{hyperref}% add hypertext capabilities
\usepackage[mathlines]{lineno}% Enable numbering of text and display math
\usepackage{tikz}
\usetikzlibrary{positioning}
\usepackage{qcircuit}
\usepackage{braket}
\usepackage{indentfirst}
\usepackage{float}
\begin{document}

%\preprint{APS/123-QED}

\title{Dimensional interpolation for metallic hydrogen}
% Force line breaks with \\
%\title{Dimensional Scaling for Quantum Criticality on a Quantum Computer}
%\maketitle
%\vspace{-2em}

%%%%%%%%%%%%%%%%%%%%%%%%%%%%%%%%%
%\author{Kumar J. B. Ghosh\textsuperscript{1} Sabre Kais \textsuperscript{2} Dudley. R. Herschbach\textsuperscript{3} \endgraf
%\scriptsize\itshape \textsuperscript{1}Department of Electrical and Computer Engineering, University of Denver, Denver, CO, 80210, USA. \\ \textsuperscript{2} Chemistry department and Physics, Purdue University, West Lafayette, IN, 47906, USA.\\ \textsuperscript{3} Department of Chemistry and Chemical Biology, Harvard University  Cambridge MA 02138, USA.}

%
\author{Kumar J. B. Ghosh}
\email{jb.ghosh@outlook.com}
\affiliation{Department of Electrical and Computer Engineering, University of Denver, Denver, CO, 80210, USA.
}%
\author{Sabre Kais}
\email{kais@purdue.edu}
\affiliation{ Department of Chemistry  and Physics,
Purdue University, West Lafayette, IN, 47906, USA.
}%
\author{Dudley R. Herschbach}
\email{dherschbach@gmail.com}
\affiliation{ Department of Chemistry and Chemical Biology, Harvard University, Cambridge MA 02138, USA 
}%

%\author{Kumar J B Ghosh, Teng Bian, Vivek Dixit, Sabre Kais}
%\email{kais@purdue.edu}
%\affiliation{%
%Chemistry department and Physics, Purdue University,\\
% West Lafayette, IN, 47906, USA.
%}%


%%%%%%%%%%%%%%%%%%%%%%%%%%%%%%%%%%%%%%%%%%%%%%%%%%%%%%%%%%%%%%%%%%%%%%%%%%%%%

%\author{Dudley. R. Herschbach}
%% %\homepage{http://www.Second.institution.edu/~Charlie.Author}
%\affiliation{
% Chemistry department, Harvard University


%\date{}% It is always \today, today,

\begin{abstract}
We employ a simple and mostly accurate dimensional interpolation formula using dimensional limits  $D=1$ and $D=\infty$ to obtain $D=3$ ground-state energy of metallic hydrogen. 
We also present results describing the phase transitions for different symmetries of  three-dimensional structure lattices.    The interpolation formula not only predicts fairly accurate energies but also predicts a correct functional form of the energy as a function of the lattice parameters.  That allows us to calculate different physical quantities such as the bulk modulus, Debye temperature, and critical transition temperature, from the gradient and the curvature of the energy curve as a function of the lattice parameters.  These theoretical calculations suggest that metallic hydrogen is a likely candidate for high temperature superconductivity.   The dimensional interpolation formula is robust and might be useful to obtain the energies of complex many-body systems.  
%\begin{description}
%\end{description}
\end{abstract}

\maketitle

%%%%%%%%%%%%%%%%%%%%%%%%%%%%%
%\tableofcontents
%\newpage
%%%%%%%%%%%%%%%%%%%%%%%%%%%

%\tableofcontents

%\doublespacing

\section{\label{sec: Introduction}Introduction}

In 1935, Eugene Wigner and H.B. Huntington predicted the metallization of hydrogen\cite{WignerandHuntington}, a phase of hydrogen that behaves like an electrical conductor. 
Ever since this has been a major quest for condensed matter physics.  In pursuing metallic hydrogen (MH), we have admired many papers, but cite a few dealing with extreme high-pressure experiments \cite{hemley1988phase, hemley1991high, hemley1996synchrotron, zaghoo2016evidence, Silvera_2018, loubeyre2020synchrotron}.  Moreover, MH is a candidate for phase transitions from superconductivity to superfluidity and vice versa under the influence of a magnetic field \cite{babaev2007violation, babaev2004superconductor, PhysRevLett.89.067001}. 

Dimensional scaling offers simple solutions for $D = 1$ and $D \to \infty$ limits, then often interpolates to obtain useful results for $D = 3$, with accuracy adequate for many areas of chemical physics \cite{herschbach2012dimensional, goodson1992large, zhen1993large, kais19941, rudnick1987shapes, loeser1991dimensional, kais1993dimensional, wei2008dimensional, wei2007dimensional, Herschbach2017, germann1993large}.    Already, the $D \to \infty$ limit for MH was treated in 1992 by John Loeser \cite{loeser1993large}.  He employed a Hartree-Fock Hamiltonian that localizes the electrons in a lattice of hydrogen atoms with clamped nuclei for rigid three-dimensional simple cubic (SC), body-centered (BCC), face-centered (FCC) cubic proton-lattices.  We shall tune up the   $D \to \infty$ limit and develop the $D = 1$ limit for MH and interpolate to obtain $D = 3$.  Recently, we used dimensional interpolation to apply ground-state energies for two, three, and four electron atoms and for ground-state H$_2$ molecule over a wide range of the internuclear distance $R$.   The results compare well with the standard exact energies obtained from the Full Configuration Interaction \cite{ghosh2020unorthodox}.

Here we investigate the phase transition for metallic hydrogen using dimensional scaling with the three-dimensional proton lattices: SC, BCC, and FCC.  Section \ref{sec:Dimensional scaling formula for extended systems} describes an interpolation formula for extended systems. Sections \ref{sec: Critical phenomena large-D-limit} and
\ref{sec: Critical phenomena in one dim} develop a one-electron Hamiltonian for quantum theory of MH in SC, BCC, and FCC lattice with $D \to \infty$ and $D = 1$ respectively. Section \ref{sec: Critical phenomena in three dim} implements the dimensional interpolation to obtain $D = 3$, observes MH and its phase transition. Section \ref{sec: elastic modulus} uses extrapolation and curve fitting to obtain a functional form of the ground state energy of MH in SC, BCC, FCC lattice as a function of the lattice parameter, $R$, and investigate the physical properties like bulk modulus, Debye temperature, melting curve, and superconductivity.  Section \ref{sec:Conclusion} has some prospects.

\section{Dimensional interpolation formula for extended systems} \label{sec:Dimensional scaling formula for extended systems}

For dimensional scaling of atoms and molecules the energy erupts to infinity as $D\to 1$ and vanishes as $D\to \infty$. Hence, we adopt scaled units (with hartree atomic units) whereby $E_D = \left( Z/\beta \right)^2 \epsilon_D$ and $\beta = \frac{1}{2} \left(D - 1 \right)$, so the reduced energy $\epsilon_D$ remains finite in both limits. When expressed in a $1/Z$ perturbation expansion, the reduced energy is given by
\begin{equation}
\epsilon_D = -1 +  \epsilon_D^{(1)} \lambda+   \epsilon_D^{(2)} \lambda^2 + ~... ~  \label{perturbation}
\end{equation}
with $\lambda = 1/Z$, where $Z$ is the total nuclear charge of the corresponding atom.

The interpolation for atoms, developed in Ref. \citep{Herschbach2017}, weights the dimensional limits by $\delta = 1/D$, providing $\delta \epsilon_1$ and $\left( 1 - \delta \right) \epsilon_\infty$ in a simple analytic formula
\begin{equation}
\epsilon_D = \delta \epsilon_1 + \left( 1 - \delta \right) \epsilon_\infty + \left[ \epsilon_D ^{(1)} - \delta \epsilon_1^{(1)} - \left( 1 - \delta \right) \epsilon_\infty^{(1)} \right] \lambda . \label{interpolation_formula}
\end{equation}

For a diatomic molecule, a different scaling scheme is used and illustrated. The rescaling of distances is:

\begin{equation}
R \to \delta R^\prime \text{ for } D \to 1 ; ~ R \to \left( 1 - \delta \right) R^\prime \text{ for } D \to \infty . \label{scalingforinterpolation}
\end{equation}

An approximation for $D = 3$ (where $R = R^\prime$) emerges:

\begin{equation}
\epsilon_3 \left( R^\prime \right) = \frac{1}{3} \epsilon_1 \left( \frac{1}{3} R^\prime \right) + \frac{2}{3} \epsilon_\infty \left( \frac{2}{3} R^\prime \right), \label{scaledinterpolationformula}
\end{equation}
interpolating linearly between the dimensional limits  \cite{frantz1988, Tan_and_Loeser, lopez1993scaling, loeser1994correlated}.

We keep the structure of the formula same as above for extended systems like metallic hydrogen. We assume the cubic symmetry for the metallic hydrogen in the large D limit and a linear chain of atoms in one dimension. The rescaled distances in different dimension is given by:

\begin{subequations}\label{rescaled_distance}
\begin{eqnarray}
\text{In } D = 1: ~ r \to r^\prime/3 \text{ and } R \to R^\prime/3   \,;
\end{eqnarray}
\begin{eqnarray}
\text{In } D \to \infty: ~ \rho \to 2 \rho^\prime/3  \text{ and } R \to 2 R^\prime/3   \,.
\end{eqnarray}
\end{subequations}

The coordinates $r$ and $\rho$ are the electronic coordinates, and the parameters $R$ is the spacing between the atomic nuclei (or lattice parameters) in $D=1$ and $D=\infty$ respectively.

\section{\label{sec: Critical phenomena large-D-limit} Metallic hydrogen at the large-D-limit}
%Dimensional scaling as applied to MH was examined by John Loeser \cite{loeser1993large}. 
With appropriate scaling, energies will be in units of $4/(D-1)^2$ hartrees, and distances in units of $D(D–1)/6$ bohr radii. Loeser applied with simplifications the Hartree-Fock one-electron Hamiltonian in the $D \to \infty$ limit in a lattice of hydrogen atoms with clamped nuclei:   

\begin{equation}
\mathcal{H} = \frac{9}{8 \rho^2} -  \frac{3}{2 \rho} + W(\rho,R), \label{Hamiltonianmetallichydrogen}
\end{equation}
where
\begin{equation}
W(\rho,R) = \frac{3}{4} \sum_{l,m,n \in \mathcal{L}^\prime}\frac{1}{\sqrt{\sigma^2R^2}} -\frac{2}{\sqrt{\sigma^2R^2 + \rho^2}}+\frac{1}{\sqrt{\sigma^2R^2 + 2\rho^2}},
\end{equation}
with
\begin{equation}
\sigma^2 = l^2+m^2+n^2.
\end{equation}

For any specified lattice type and scaled lattice constant $R$, the minimum of Eq. (\ref{Hamiltonianmetallichydrogen}) with respect to $\rho$ gives the energy per electron.  The whole lattice is three-dimensional, noted $\mathcal{L}^\prime$ minus the one site $(0,0,0)$.  The single variable $\rho$ is the orbit radius and $R$ is the lattice spacing. The quantity $(\rho / R)$ is used to characterize the different density regimes.

We began with the simple cubic lattice.  In Hartree-Fock approximation of MH the triple sum $W(\rho, R)$ is a kind of Madelung sum.  Loeser evaluated $W(\rho, R)$ for the two limiting cases.    For, the low-density regime ($\rho\ll R$), the sums over integers are taken from the tables of Born and Misra \cite{born1940stability} and the Hamiltonian of Eq. (\ref{Hamiltonianmetallichydrogen}) becomes: 

\begin{equation}
\lim_{\rho\ll R} \mathcal{H} = \frac{9}{8 \rho^2} -  \frac{3}{2 \rho} + 3.89157 \left(\frac{3 \rho^4}{2 R^5}\right). \label{metalliclowdensity}
\end{equation}

For the high-density regime $(\rho\gg R)$, the sum can be replaced by an integral and the Hamiltonian of Eq.  (\ref{Hamiltonianmetallichydrogen}) becomes:

%and for high density limit $(\rho\gg R)$ the above Hamiltonian (\ref{Hamiltonianmetallichydrogen}) boils down to
\begin{equation}
\lim_{\rho \gg R}\mathcal{H} = \frac{9}{8 \rho^2} -  \frac{3}{2 \rho} + 2.17759 \left(\frac{3 \rho^2}{2 R^3}\right). \label{metallichighdensity}
\end{equation}

\begin{figure}[H]
\centering
  \begin{subfigure}{0.5\textwidth}
    \includegraphics[width=\linewidth]{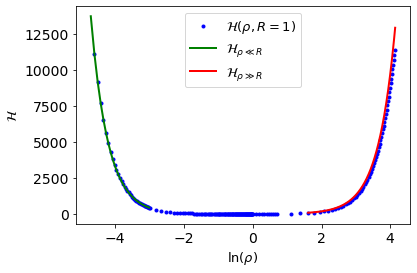}
    \caption{} \label{fig:1a}
  \end{subfigure}%
  \hspace*{\fill}   % maximize separation between the subfigures
  \begin{subfigure}{0.5\textwidth}
    \includegraphics[width=\linewidth]{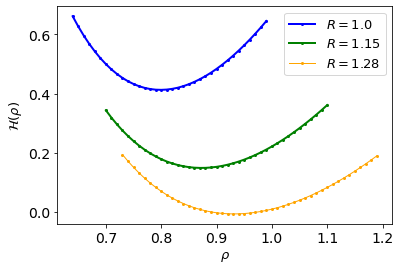}
    \caption{} \label{fig:1b}
  \end{subfigure}%
  
\caption{At the left \ref{fig:1a}, we plot the electronic energy per atom  $H(\rho, R)$, for MH in SC lattice, as a function of $\ln (\rho)$ keeping the parameter $R = 1$. We also plot the energy for $\rho \gg R$ obtained from Eq.(\ref{metallichighdensity}) in red, and the energy for $\rho \ll R$ obtained from Eq. (\ref{metalliclowdensity}) in green. At right \ref{fig:1b} we plot the energy for each atom $\mathcal{H} (\rho)$ as a function of $\rho$ for $R= 1.0, 1.5$ and $1.28$ respectively.}
\label{metallichydrogen}
\end{figure}

For the criticality of the metallic hydrogen, we have to find the critical density for which the above Hamiltonian (\ref{Hamiltonianmetallichydrogen}) attends the minimum. For calculation-convenience we fix the parameter $R = 1$ and the lattice size to be $(400\times400\times400)$. Although ideally the number lattice points should be infinitely large, but for calculation  our lattice size is sufficient to observe the critical phenomenon. We calculate numerically and display in Fig.\ref{fig:1a} of the above Hamiltonian for a wide range of $\rho$, from 0.01 to 65, i.e. the density $(\rho/R)$ ranges from 0.01 to 65.

From Fig. \ref{fig:1b},  we find that, for $R=1$, at density $\rho/R = 0.799$ Hamiltonian (\ref{Hamiltonianmetallichydrogen}) attends the minimum. 
%Although this a very simple model at $D \to \infty$ limit with Hartree-Fock approximation and hydrogen atoms are in a simple cubic lattice, but we observe the criticality and find a critical density for the corresponding critical phenomenon.
Then we choose different values of $R$ and calculate the minimum of the Hamiltonian $\mathcal{H} (\rho^\star , R)$ by varying the parameter $\rho$ for each $R$, which gives the electronic ground state energy per atom. 
 
\begin{figure}[H]
\centering
  \begin{subfigure}{0.5\textwidth}
    \includegraphics[width=\linewidth]{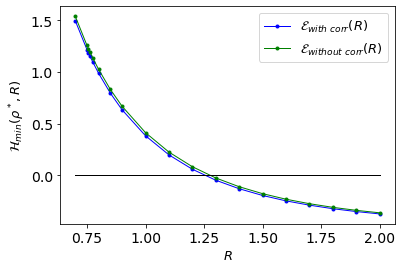}
    \caption{} \label{fig:2a}
  \end{subfigure}%
  \hspace*{\fill}   % maximize separation between the subfigures
  \begin{subfigure}{0.5\textwidth}
    \includegraphics[width=\linewidth]{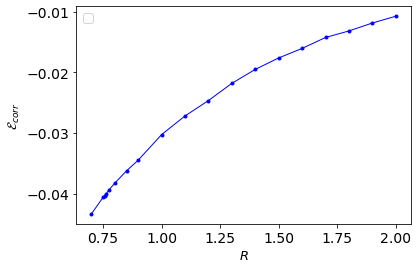}
    \caption{} \label{fig:2b}
  \end{subfigure}%
  
\caption{At the left \ref{fig:2a}, we plot the energy obtained at $D \to \infty$ with inter-electronic correlation in blue, compared with the HF energy as a function of $R$ in green. At the right side Fig. \ref{fig:2b}, the difference between the correlation energy and the HF energy $\mathcal{E}_{corr}$ is plotted as a  function of $R$.}
\label{correlation_energy_inf}
\end{figure}

At $R = 1.28$, we find $\rho^\star = 0.932$ at the minimum and see in Fig. \ref{fig:2a}, the ground state energy becomes positive to negative around $R = 1.28$.  Physically this means for $R < 1.28$, the ground state energy becomes positive therefore makes the system is unstable. Therefore, at the point $R = 1.28$ and $\rho^\star = 0.932$, the system goes through a phase transition.   This is an elemental model at $D \to \infty$ limit with Hartree-Fock approximation and hydrogen atoms are in a simple cubic lattice, but we observe the criticality and find a critical density for the corresponding critical phenomenon. 

Remarkably, Loeser added to the Hartree-Fock approximation in the $D \to \infty$ limit by introducing inter-electronic correlation.  That was essentially by opening the dihedral angles in the simple cubic lattice \cite{loeser1993large}.   As shown in Fig. \ref{fig:2a}, the energy with correlation is nearly the same as without the correlation.    Further, in Fig. \ref{fig:2b}, we plot the difference between the correlation energy and the Hartree-Fock energy. 

\begin{figure}[H]
\centering
  \begin{subfigure}{0.5\textwidth}
    \includegraphics[width=\linewidth]{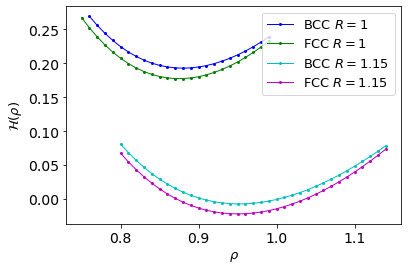}
    \caption{} \label{fig:3a}
  \end{subfigure}%
  \hspace*{\fill}   % maximize separation between the subfigures
  \begin{subfigure}{0.5\textwidth}
    \includegraphics[width=\linewidth]{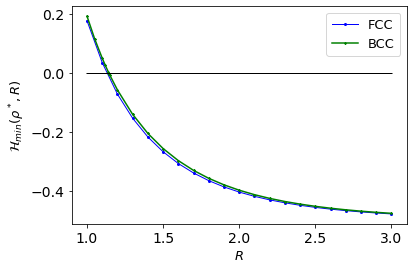}
    \caption{} \label{fig:3b}
  \end{subfigure}%
  
\caption{At left \ref{fig:3a} we plot the energy per atom $H(\rho, R)$, for MH in BCC and FCC lattices, as a function of the parameter $\rho$  fixing $R = 1$ and $1.15$ respectively.  At right \ref{fig:3b}, we plot the ground state energy $\mathcal{H}_{min} (\rho^\star , R)$ as a function of $R$.}
\label{metallichydrogen2}
\end{figure}

We first choose the parameter $R = 1$ in the electronic Hamiltonian per atom and calculate the energies with respect to the variable $\rho$; see Fig \ref{fig:3a} at left.    Like the SC case we find a minimum for the Hamiltonian, which is $\rho_{BCC} = 0.88 = \rho_{FCC}$.

Again we calculate the minimum of the Hamiltonian $\mathcal{H} (\rho^\star , R)$ for the different values of $R$ by varying the parameter $\rho$, which gives the electronic ground state energy per atom.  From Fig. \ref{fig:3b} at right , we see that the ground state energy becomes positive to negative around $R = 1.15$ for both FCC and BCC lattices. Hence, the ground state energy becomes positive so makes the lattice is unstable or the system loses the crystalline structure. Therefore, around the point $R_{BCC} = 1.15 =  R_{FCC}$   and the system goes through a phase transition.  Since the parameters ($\rho , R$) at the transition point are same for BCC and FCC lattices, this could be coming from some symmetry working akin to their reciprocal spaces.

\section{\label{sec: Critical phenomena in one dim}Metallic hydrogen in $D=1$} 

A lonely, single hydrogen molecule in one-dimension is described by I. R. Lapidus \cite{Lapidus1975, Lapidus1982}.  However, to investigate the quantum theory of MH in one dimension, we need to develop a long chain of $N$ individual hydrogen atoms, with the nuclei sitting on the lattice sites located at $r= 0, R, 2R, ... (N-1)R$.  Hence one-electron Hamiltonian in $D = 1$ using atomic units can be written as:

\begin{equation}
\mathcal{H} =  -\frac{1}{2} \frac{\partial^2}{\partial r^2} -  \sum_{i=0}^{N-1} \delta (r+ iR) + \text{electron-electron interaction part}. \label{Hamiltonianhydrogen1D}
\end{equation}

We choose the linear combination of atomic orbitals (LCAO) representation and construct the state for each electron in the metallic hydrogen chain in one dimension as follows:

\begin{equation}
\psi (r) = \frac{1}{\mathcal{N}} \left[ \phi_0 (r) + \phi_1 (r) + \phi_2 (r) + ... \right],
\end{equation}
\noindent
with the of the normalization constant 
\begin{equation}
\mathcal{N} = \left[\sum_{i,j} e^{-|i-j| R}  \left( 1 + |i-j| R \right) \right]^{1/2},
\end{equation}
\noindent
and the individual normalized wave functions
$$\phi_i (r)=  e^{-|r +iR|}. $$

The kinetic and potential energy part of the above Hamiltonian is calculated as:
\begin{eqnarray}
E_{KE+PE}&=& \big \langle \psi (r) \big| -\frac{1}{2} \frac{\partial^2}{\partial r^2} -  \sum_{k=0}^{N-1} \delta (r+ kR)~ \big| \psi (r) \big \rangle \nonumber \\
&=& \frac{1}{\mathcal{N}^2} \left[ -\frac{1}{2} \sum_{i,j}  \left( -1 + |i-j| R \right) e^{-|i-j| R}  - \sum_{i,j,k} e^{-|i-k| R}e^{-|j-k| R} \right].
\end{eqnarray}

Unlike the nuclei, which are localized at single points, the superpositioned electron-clouds are smeared around  the whole lattice. So, to calculate the inter-electronic repulsion part we consider an electron density (or a negative charge density) over an infinitesimally small line element $dr$ at a location $r$ which interacts with another electron density sitting over an infinitesimally small line element $dr^\prime$ at a location $r^\prime$. The repulsion energy for these two infinitesimally small electron clouds is given by $d E_{ee} = \psi (r) \psi (r) \delta \left( r - r^\prime \right)  \psi (r^\prime) \psi (r^\prime)~ dr dr^\prime$. Therefore, the total electron-electron interaction part is given by:

\begin{eqnarray}
E_{ee}&=& \int_{\infty}^\infty \int_{\infty}^\infty d r dr^\prime~ \psi (r) \psi (r^\prime) \delta \left( r - r^\prime \right) \psi (r) \psi (r^\prime) \nonumber \\
&=& \frac{1}{\mathcal{N}^4} \int_{\infty}^\infty d r \sum_{i,j,k,l} \phi_i (r) \phi_j (r) \phi_k (r) \phi_l (r) \nonumber \\
&=& \frac{1}{\mathcal{N}^4} \int_{\infty}^\infty d r \sum_{i,j,k,l} e^{-|r-ir|}e^{-|r-jR|}e^{-|r-kR|}e^{-|r-lR| }. \label{interaction}
\end{eqnarray}

We break the above expression  (\ref{interaction}) into many smaller terms and calculate all them individually as follows:
$$
\int_{\infty}^\infty d r ~ \phi^4 (r) = \frac{1}{2},
$$

$$
\int_{\infty}^\infty d r ~ \phi_i^2 (r) \phi_j^2 (r) = e^{-2|i-j| R} \left( \frac{1}{2} + |i-j| R \right), \text{ for } i\neq j,
$$

$$
\int_{\infty}^\infty d r ~ \phi_i (r) \phi_j^3 (r) = \frac{3}{4} e^{-|i-j| R} - \frac{1}{4} e^{-3|i-j| R}, \text{ for } i\neq j,
$$

$$
\int_{\infty}^\infty d r ~ \phi_i (r) \phi_j^2 (r) \phi_k (r) = \frac{1}{2} e^{(k-i) R} - \frac{1}{4} e^{(3k-i-2j) R} - \frac{1}{4} e^{(2j+k-3i) R}, \text{ for } i> j>k,
$$

$$
\int_{\infty}^\infty d r ~ \phi_i (r) \phi_j^2 (r) \phi_k (r) =  e^{(i+k-2j) R} \left[\frac{3}{4} + (j-i)R\right] - \frac{1}{4} e^{(3k-i-2j) R}, \text{ for } j> i>k,
$$

$$
\int_{\infty}^\infty d r ~ \phi_i (r) \phi_j^2 (r) \phi_k (r) =  e^{(2j-k-i) R} \left[\frac{3}{4} + (k-j)R\right] - \frac{1}{4} e^{(2j+k-3i) R}, \text{ for } i> k>j,
$$

$$
\int_{\infty}^\infty d r ~ \phi_i (r) \phi_j (r) \phi_k (r) \phi_l (r) =  e^{(k+l-i-j) R} \left[1 + (k+j)R\right] - \frac{1}{4} e^{(j+k+l-3i) R}- \frac{1}{4} e^{(3l-i-j-k) R}, \text{ for } i> k>j>l,
$$ 

and we use the multinomial theorem:

$$
\left( x_1 + x_2+ ... x_m \right)^n = \sum_{k_1+k_2+...k_m =n} \frac{n!}{k_1! k_2!...k_m!}  \prod_{t=1}^m (x_t)^{k_t}
$$

to calculate the interaction part in equation (\ref{interaction}).

We set the number of lattice points equal to $i, j, k, l = 100$ and calculate all the above quantities numerically and obtain the final energy as a function of the inter-atomic distance $R$.

\begin{figure}[H]
    \centering
    \begin{tikzpicture}[
 image/.style = {text width=0.7\textwidth, 
                 inner sep=0pt, outer sep=0pt},
node distance = 1mm and 1mm
                        ] 
\node [image] (frame1)
    {\includegraphics[width=\linewidth]{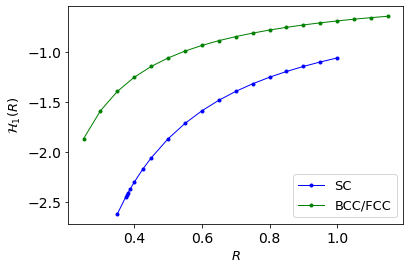}};
%\node [image,right=of frame1] (frame2) 
%    {\includegraphics[width=\linewidth]{metallic_hydrogen/one_dim_energy_vs_R_fcc}};
\end{tikzpicture}
\caption{We plot the energy $\mathcal{H}_1 (R)$ for MH at $D=1$  as a function of $R$; blue curve is for SC lattice and green for FCC or BCC lattice.}
\label{metallichydrogen1dim}
\end{figure}

For FCC or BCC lattices the calculations in $D=1$ is slightly different from SC lattice in one dimension. For e.g. in case of BCC lattice in $D=\infty$ the Hamiltonian (\ref{Hamiltonianmetallichydrogen}), the lattice parameters $l, m, n$ are either even or odd integers \cite{loeser1993large}. We choose the same conversion for the calculation in one dimension. In one dimension  the FCC and BCC lattices are the same.

For MH in $D=1$ with inter-electronic correlation there is no way to define dihedral angles between electrons. Therefore, we keep the Hamiltonian same as defined in Eq. (\ref{Hamiltonianhydrogen1D}). In Fig. \ref{metallichydrogen1dim}, we plot the energy per electron as a function of the inter-atomic distance $R$ for simple cubic and FCC/BCC lattices.

\section{\label{sec: Critical phenomena in three dim} Metallic hydrogen in $D=3$ from dimensional interpolation}

The dimensional interpolation formula described in section \ref{sec:Dimensional scaling formula for extended systems} combines the $D=1$ and $D=\infty$ limits and obtains the $D = 3$ reduced energy from  

\begin{equation}
\epsilon_3 \left( R \right) = \frac{1}{3} \epsilon_1 \left( \frac{1}{3} R \right) + \frac{2}{3} \epsilon_\infty \left( \frac{2}{3} R \right). \label{equation16}
\end{equation}

\begin{figure}[H]
    \centering
    \begin{tikzpicture}[
 image/.style = {text width=0.7\textwidth, 
                 inner sep=0pt, outer sep=0pt},
node distance = 1mm and 1mm
                        ] 
\node [image] (frame1)
    {\includegraphics[width=\linewidth]{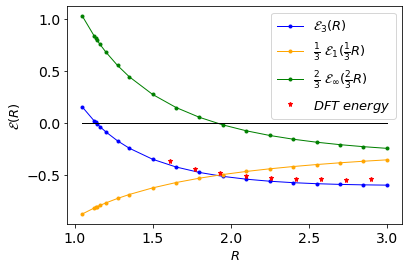}};
%\node [image,right=of frame1] (frame2) 
%    {\includegraphics[width=\linewidth]{metallic_hydrogen/Correlation_interpolation}};
\end{tikzpicture}
\caption{The orange and green curves describe the results from $D=1$ and $D \to \infty$  respectively. The blue curve energy $\mathcal{E}_3 (R)$ represented the interpolation result at $D = 3$ for MH, in the SC lattice.  The red points represent the results obtained from density functional theory in \cite{PhysRevB.30.5076}. The black line shows the zero-energy line, where the blue curve goes through on $R = 1.14$ at a phase transition.}
\label{metallichydrogen3dim}
\end{figure}

In Fig. \ref{metallichydrogen3dim}, we plot the electronic energies involved for the metallic hydrogen in simple cubic lattice (SC). We compare our interpolation $\mathcal{E}_3 (R)$ curve (blue) with points (red) that come from density functional theory \cite{PhysRevB.30.5076} (there Table III).   The agreement is very good.    As noted in Fig. \ref{correlation_energy_inf}, the inter-electronic correlation examined in the $D \to \infty$ limit turns quite minor.    From Fig. \ref{metallichydrogen3dim} we see that the ground state energy at $D = 3$ becomes positive to negative around $R = 1.14$.  Physically this means for $R < 1.14$ the ground state energy becomes positive therefore makes the lattice is unstable or the system loses the crystalline structure. Therefore, at the point $R = 1.14$, the system goes through a phase transition.   

\begin{figure}[H]
\centering
  \begin{subfigure}{0.5\textwidth}
    \includegraphics[width=\linewidth]{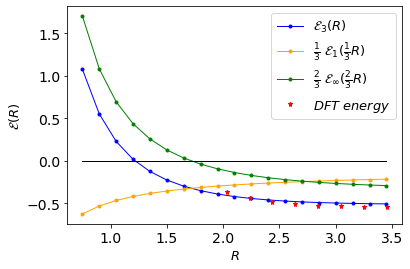}
    \caption{} \label{fig:6a}
  \end{subfigure}%
  \hspace*{\fill}   % maximize separation between the subfigures
  \begin{subfigure}{0.5\textwidth}
    \includegraphics[width=\linewidth]{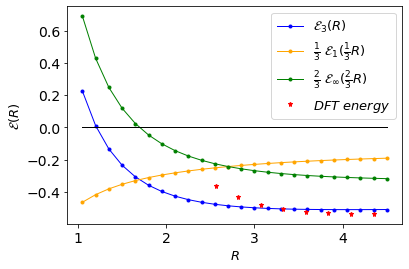}
    \caption{} \label{fig:6b}
  \end{subfigure}%
  
\caption{The panel left \ref{fig:6a} is for the  BCC  lattice, the panel right  \ref{fig:6b} for the FCC lattice.  Compare results in Fig. \ref{metallichydrogen3dim} for the SC lattice.    Both of the blue curves meet the black line at zero-energy on $R= 1.14$ at the same phase transition for all three lattices}
\label{interppolation_bcc}
\end{figure}

%%%%%%%%%%%%%%%%%%%%%%%%%%%%%%%%%%%%%%%%%%%%%%%%%%%%%%%%%%%%%%%%
Fig. \ref{interppolation_bcc} displays the MH energies from the BCC and FCC lattices, similar to the SC lattice in Fig. \ref{metallichydrogen3dim}, interpolating to $D = 3$ from Eq.(\ref{equation16}). The corresponding transition point for BCC and FCC lattice, where the ground state energy changes sign, is given by $R_{FCC} = 1.21 = R_{BCC}$.

%%%%%%%%%%%%%%%%%%%%%%%%%%%%%%%%%%%%%%%%%%%%

\section{\label{sec: elastic modulus}Physical properties of Metallic hydrogen}

The scaled lattice constant $R$ is related to $r_s$, the standard solid state parameter defined as the radius of a sphere (in $a_0$  bohr units) in which contains on average one electron.    For the SC lattice, 
\begin{equation}
\frac{4}{3} \pi r_s^3 = R^3 . \label{rs_simple_cubic}
\end{equation}
Thus, $r_s = 0.71$ corresponds to the transition point $R = 1.14$, for the $D \to \infty$ limit. Many studies for $D = 3$ have obtained $r_s = 0.8$ for the existence of crystalline phase of metallic hydrogen \cite{PhysRevB.21.2641}. However, we have simply used $R$ when the interpolation formula provides appropriate energies of metallic hydrogen for different lattice symmetries and parameters. 

\begin{figure}[H]
    \centering
    \begin{tikzpicture}[
 image/.style = {text width=0.7\textwidth, 
                 inner sep=0pt, outer sep=0pt},
node distance = 1mm and 1mm
                        ] 
%\node [image] (frame1)
%    {\includegraphics[width=\linewidth]{metallic_hydrogen/BCC/comparison}};
\node [image] (frame1)
    {\includegraphics[width=\linewidth]{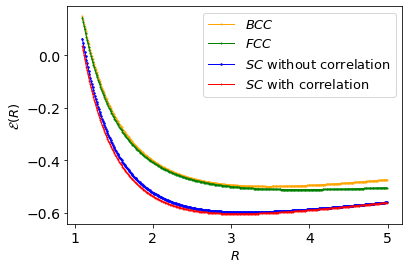}};
\end{tikzpicture}
\caption{The simple formula, Eq. (\ref{curve_fit_energy}), fits neatly the energies of MH in Table \ref{table:lattice parameters}: SC with HF (in blue) and with correlation (in red); BCC (in orange), and FCC (in green) lattice structures respectively.}
\label{energy compare bcc sc} 
\end{figure}

In this section, we are allowed to calculate different physical quantities such as the bulk modulus, Debye temperature, and critical transition temperature, from the gradient and the curvature of the energy curve as a function of the lattice parameters.   Thus, the numerical results of the interpolation formula can be fitted to the following functional form as a function of the lattice parameter $R$  \cite{PhysRev.111.442, PhysRev.158.876}, driven in a following table and Fig. \ref{energy compare bcc sc}.

%So far, we  have shown that the interpolation formula gives very accurate energies of metallic hydrogen for different lattice symmetries and parameters. In this section, we will show that the interpolation formula also predicts  the correct functional form of the energy  as a function of the lattice parameters and symmetries.  Thus, allowing us to calculate different physical quantities such as the bulk modulus, Debye temperature, and critical transition temperature,  from the gradient and the curvature of the energy curve  as a function of the lattice parameters. 
%
%The numerical results of the interpolation formula can be fitted to the following functional form as a function of the lattice parameter $R$  \cite{PhysRev.111.442, PhysRev.158.876}, 

\begin{equation}
\mathcal{E} (R) = \frac{A}{R^2} + \frac{B}{R}+ C- D \ln (R), \label{curve_fit_energy}
\end{equation} 
\noindent
with the parameters given in the following table \ref{table:lattice parameters} :

\begin{table}[H]
\caption{Parameters describing $\mathcal{E} (R)$ in different lattice structures} % title of Table
\centering % used for centering table
\begin{tabular}{| c | c | c | c | c |} % centered columns (4 columns)
\hline %inserts double horizontal lines
Lattice structure & ~A~ & ~B ~&~ C~ &~ D ~\\ [0.5ex] % inserts table
%heading
\hline % inserts single horizontal line
SC in HF limit ~&~  1.3236  ~&~  -0.0975  ~&~ -0.9650  ~&~  -0.2311 \\ % inserting body of the table
SC with correlation\textit{$^{a}$} ~&~ 1.3458 ~&~ -0.1689  ~&~ -0.9457 ~&~ -0.2264 \\ %
BCC ~&~  0.7889 ~&~  0.7273  ~&~  -1.1974  ~&~  -0.3397 \\ %
FCC ~&~  1.1628 ~&~  -0.1070  ~&~  -0.7329  ~&~  -0.1265 \\ [1ex]
%5 & 45 & 300 & 556 \\ [1ex] % [1ex] adds vertical space
\hline  %inserts single line

\end{tabular}
\label{table:lattice parameters} % is used to refer this table in the text
\newline \newline { \textit{$^{a}$}  See Fig. \ref{correlation_energy_inf}  has added SC with inter-electronic correlation. Others with Hartree-Fock (HF).}
\end{table}

We also calculate the $\mathcal{E}_{min}$ and $R_{min}$ from  Eq. (\ref{curve_fit_energy}) and in Table \ref{table:energy minimum} compare them for the lattice structures:

\begin{table}[H]
\caption{$\mathcal{E}_{min}$ and $R_{mim}$ in different lattice structures} % title of Table
\centering % used for centering table
\begin{tabular}{| c | c | c |} % centered columns (4 columns)
\hline %inserts double horizontal lines
Lattice structure & $R_{mim}$ & $\mathcal{E}_{min}$  \\ [0.5ex] % inserts table
%heading
\hline % inserts single horizontal line
SC in HF limit ~&~  3.18  ~&~  -0.597  \\ % inserting body of the table
SC with correlation ~&~ 3.09 ~&~ -0.603  \\ %
BCC ~&~  3.48 ~&~  -0.499   \\ %
FCC ~&~  3.88 ~&~  -0.512 \\ [1ex]
%5 & 45 & 300 & 556 \\ [1ex] % [1ex] adds vertical space
\hline %inserts single line
\end{tabular}
\label{table:energy minimum} % is used to refer this table in the text
\end{table}

The minimum energies of the MH lattices are evident in Fig. \ref{energy compare bcc sc} and the energy differences are $\Delta_{FCC-SC}= 0.091$, $\Delta_{BCC-SC}= 0.104$, $\Delta_{FCC-BCC}= 0.013$  respectively.   The energy differences between FCC and BCC are very modest compared to SC.   That may lead to  possibility for a phase transition from BCC to FCC structure or vice-versa. 

Now we examine briefly consequential properties that involve from the interpolated formula (Eq. \ref{equation16}) for MH energy.   First is pressure:
 
\begin{equation}
P = -\frac{d\mathcal{E}}{dV}= -\frac{\eta}{3R^{2}}  \frac{d\mathcal{E}}{dR},
\label{pressure_formula}
\end{equation}
\noindent
with $\eta =$ number of atoms in a unit cell. And we see from the graph that around $R=3.18$, where $\mathcal{E}=\mathcal{E}_{min}$, the pressure changes sign; the corresponding $r_s = 1.9$. This transition has a physical significance. Although at high densities the crystalline phase is preferred for metallic hydrogen, however at low densities ($r_s > 1.6$) the metallic hydrogen behaves like a fluid or liquid metal \cite{PhysRevB.21.2641}, \cite{Robitaille}, \cite{bonev2004quantum}. This particular point $R=3.18$ and $\mathcal{E}= -0.597$ possibly signifies that crystalline-fluid transition point for MH in simple cubic lattice. The corresponding points for BCC and FCC lattices are $R_{BCC}=3.48$ and $R_{FCC}=3.88$ respectively.

\begin{figure}[H]
    \centering
    \begin{tikzpicture}[
 image/.style = {text width=0.7\textwidth, 
                 inner sep=0pt, outer sep=0pt},
node distance = 1mm and 1mm
                        ] 
\node [image] (frame1)
    {\includegraphics[width=\linewidth]{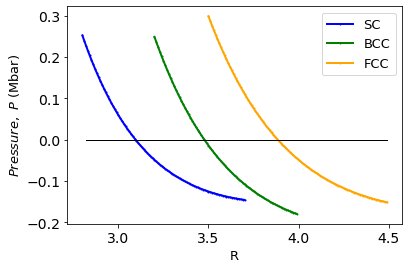}};
\end{tikzpicture}
\caption{We plot the pressure $P$ in Mbar for MH in SC (in blue), BCC (in green), and FCC (in orange) lattices as a function of $R$.}
\label{pressure}
\end{figure}
%Note that the pressure is expressed in atomic unit $E_h/a_0^3 = 294$ Mbar. 

%%%%%%%%%%%%%%%%%%%%%%%%%%%%%%%%%%%%%%%%%%%%%%%%%%%%%%%%%%%%%%%%%%%%%

From the curvature of the energy curves in Fig. \ref{energy compare bcc sc}, we can calculate the elastic modulus of MH  for  different structures as the  bulk modulus $B$ of a cubic metallic lattice is given by

%\begin{equation}
%B = -\frac{dP}{V dV}=\frac{\eta}{R} \frac{d^2 \mathcal{E}}{dR^2}, \label{bulk_modulus_MH}
%\end{equation}
\begin{equation}
B =-V \frac{dP}{dV}, \label{bulk_modulus_MH}
\end{equation} 
\noindent
with pressure, $P$, and volume of the unit cell $V = R^3$.  The shear modulus $G$ can be calculated from the same equation  \cite{phani2008relations}.    Another important quantity that is available from the MH energy via pressure is the Debye temperature $\Theta$; the formula \cite{ANDERSON1963909, LI2012197} involves quite a few items:

\begin{equation}
\Theta = \frac{h}{k} \left( \frac{3 N_A \rho}{4 \pi M} \right)^{1/3} v_m, \label{debye}
\end{equation}
\noindent
where $h$ is the Planck constant, $k$ is the Boltzmann constant, $N_A$ is the Avogadro constant, $M$ is the atomic weight, and $\rho$ is the density, and includes $B$, the bulk modulus as well $G$, the shear modulus, and

\begin{equation}
v_m = \left[\frac{1}{3} \left( \frac{2}{v_s^3} + \frac{1}{v_p^3} \right) \right]^{-1/3}; ~ 
v_s = \sqrt{\frac{G}{\rho}} ; ~ v_p = \sqrt{\frac{B + 4G/3}{\rho}}.
\end{equation}

\begin{figure}[H]
    \centering
    \begin{tikzpicture}[
 image/.style = {text width=0.5\textwidth, 
                 inner sep=0pt, outer sep=0pt},
node distance = 1mm and 1mm
                        ] 
\node [image] (frame1)
    {\includegraphics[width=\linewidth]{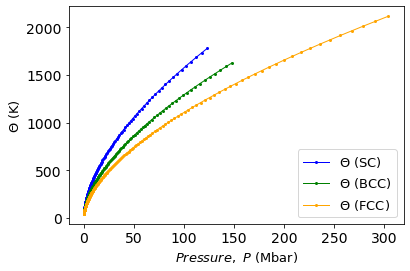}};
\node [image,right=of frame1] (frame2) 
    {\includegraphics[width=\linewidth]{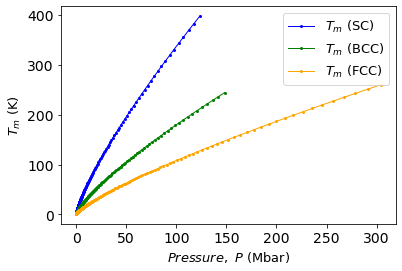}};
\end{tikzpicture}
\caption{We plot the Debye temperatures $\Theta$ (in left) and melting points $T_m$ (in right) of MH in SC, BCC, and FCC lattice structure with the correlation with respect to the pressure, $P$ .}
\label{debye_sc}
\end{figure}

Fig. \ref{debye_sc} displays Debye temperature $\Theta$ (left) and melting temperature $T_m$ (right) as functions of pressure for MH lattices, SC, BCC, and FCC.    As well known, MH will climb to a very high Debye temperature, and could become a quantum liquid.   Melting curves of metallic hydrogen as a function of pressure via the Lindemann melting law \cite{ross1977molecular, trubitsyn1971phase}.    Some samples: For SC lattice at $R = 3.1$, $\Theta = 57 K$, and at $R = 1.12$ obtains $\Theta = 1779 K$ at $P = 118$ Mbar.  For BCC lattice at $R = 3.48$, $\Theta = 52 K$, and at $R = 1.21$ obtains $\Theta = 1627 K$ at $P = 147$ Mbar.   For FCC lattice at $R = 3.88$, $\Theta = 41 K$, whereas at $R = 1.21$ obtains $\Theta = 2114 K$ at $P = 294$ Mbar.

%The prediction results are shown in Figure \ref{debye_sc}.
%
%As it is well known, metallic hydrogen has a very high Debye temperature and could be a quantum liquid. Several attempts were made to calculate the melting curve of metallic hydrogen.   Rf. \cite{ross1977molecular, trubitsyn1971phase} used the Lindemann law to calculate the melting temperature as a function of pressure, our predictions are shown in Figure \ref{debye_sc}.  For SC lattice at $R = 3.1$ the Debye temperature $\Theta = 56.77 K$, and at $R = 1.12$, $P = 117.6$ Mbar, $\Theta = 1779.01 K$.  For BCC lattice at $R = 3.48$ the Debye temperature $\Theta = 52.35 K$, whereas at $R = 1.21$, $P = 147$ Mbar, and $\Theta = 1626.87 K$.   For FCC lattice at $R = 3.88$ the Debye temperature $\Theta = 41.18 K$, whereas at $R = 1.21$, $P = 294$ Mbar, and $\Theta = 2114.56 K$.

Finally, as many authors predicted that metallic hydrogen is a superconductor below some critical temperature and others even argued that it might be a superconductor up to a room temperature \cite{PhysRevLett.21.1748, cheng2020evidence}.   Following Koblischka, et al \cite{koblischka2020relation}, we have used our MH energy curves via interpolation from Eq. (\ref{equation16}) and formulate the transition temperature $T_c$ for a MH superconductor:

\begin{equation}
T_c = \frac{h^2}{(2x)^2 2 M_L n^{-2/3} \pi k } . \label{super_conducting_temp}
\end{equation}

Here again $h$ is the Planck constant; $k$ the Boltzmann constant; $x$ is the atomic distance.   A correction factor $n$ is usually taken to be $1$ for metals, and $M_L$ is taken to be equal to the mass of a proton $m_p$.
   
\begin{figure}[H]
    \centering
    \begin{tikzpicture}[
 image/.style = {text width=0.7\textwidth, 
                 inner sep=0pt, outer sep=0pt},
node distance = 1mm and 1mm
                        ] 
\node [image] (frame1)
    {\includegraphics[width=\linewidth]{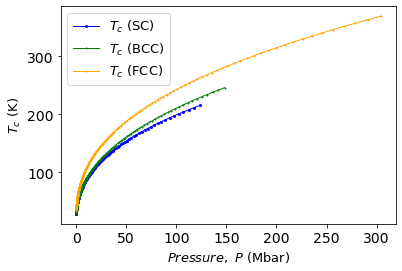}};
\end{tikzpicture}
\caption{We plot the transition temperature $T_c$ as a function of pressure, $P$.}
\label{transition temprature comparison}
\end{figure}

In Fig. \ref{transition temprature comparison}, we plot the possible values of transition temperature $T_c$ calculated from Eq. (\ref{super_conducting_temp}) as a function of pressure for MH in SC, BCC, and FCC lattices, within a range where the energy is negative and pressure is positive.  For the atomic distances, $x = R$ for SC;  $x = (\sqrt{3}/2) R$ for BCC; $x = R/\sqrt{2}$ for FCC.   Samples:   For SC lattice at $R = 3.1$, $T_c = 28 K$ whereas for $R = 1.12$, $T_c = 215 K$ and $P = 118$ Mbar.   For BCC lattice at $R = 3.48$, $T_c = 30 K$ whereas at $R = 1.21$, $T_c = 246 K$, and $P = 147$ Mbar.   For FCC lattice at $R = 3.88$, $T_c = 36 K$ whereas for $R = 1.21$, $T_c = 369 K$ and $P = 294$ Mbar.

%Here again $h$ is the Planck constant; $K_B$ the Boltzmann constant; $x$ is the atomic distance, $T_c$  the superconducting transition temperature,  $n$ is a correction factor usually taken to be $1$ for metals, and for metals $M_L$ is taken to be equal to the mass of a proton $m_p$ .
%
%We plot the possible values of $T_c$ calculated from Eq. (\ref{super_conducting_temp}) as function of pressure for MH for in SC, BCC, and FCC lattices in Fig \ref{transition temprature comparison} in a range where the energy is negative and pressure is positive.
%%In a SC lattice in a range of $R = 1.12$ to $R=3.1$, where the energy is negative and pressure is positive. 
%For SC lattice the atomic distance $x = R$. At $R= 3.1$ the transition temperature $T_c = 28 K$ whereas for $R= 1.12$, $P=117.6$ Mbar, and $T_c = 215 K$. For BCC lattice the atomic distance $x = (\sqrt{3}/2) R$. At $R= 3.48$ the transition temperature $T_c = 29.9 K$ whereas at $R= 1.21$, $P=147$ Mbar, and $T_c = 246 K$. For FCC lattice the atomic distance $x = R/\sqrt{2}$. At $R= 3.88$ the transition temperature $T_c = 36 K$ whereas for $R= 1.21$, $P=294$ Mbar, and $T_c = 369 K$.

Although the experimental verification of superconductivity in MH is yet to be confirmed, but from the above theoretical calculations we see that the metallic hydrogen is a very good candidate for high temperature superconductor.

\section{\label{sec:Conclusion}Conclusion}

The simplicity of the $D \to \infty$ limit causes the disappearance of derivatives from a Hamiltonian as well as $h \to 0$, so is a true classical limit \cite{loeser1994correlated}.   It is different from a semiclassical approximation such as WKB theory for small $h$.   The simplicity of the $D = 1$ limit keeps derivatives in a Hamiltonian and is a true hyperquantum limit.  Combining these extreme partner limits delivers the dimensional interpolation formula.   It was tried out with two-electron atoms \cite{Herschbach2017} and generalized out with few electron atoms and simple diatomic molecules \cite{ghosh2020unorthodox}.   Here we find the interpolation approach is appropriate for metallic hydrogen.   With beginning by Loeser \cite{loeser1993large}, we find the interpolation not only provided adequate energies but also the correct function forms of symmetry and lattice parameters.  From the gradient and the curvature of the energy curves as a function of the lattice parameter R, we were able to calculate some important physical quantities.  Among are the bulk and shear moduli, and three temperatures governed by pressure: the Debye temperature, the Lindemann melting temperature, and the critical transition temperature for superconductivity.
   
It is relatively easy to calculate the $D \to \infty$ and $D = 1$ limits, so the interpolation formula can predict results for the physical dimension, $D = 3$.     Therefore, $D-$scaling might approach the electronic structure of strongly correlated systems, where often traditional approaches are faced with computational difficulties!

\section*{Acknowledgements}
S. K. would like to acknowledge funding by the U.S. Department of Energy (Office of Basic Energy Sciences) under Award No. DE-SC0019215.

\bibliography{main}

\end{document}